\documentclass{article}

\usepackage[english]{babel}

\usepackage[letterpaper,top=2cm,bottom=2cm,left=3cm,right=3cm,marginparwidth=1.75cm]{geometry}

\usepackage{amsmath}
\usepackage{graphicx}% http://ctan.org/pkg/graphicx
\usepackage{array}% http://ctan.org/pkg/array
\usepackage{longtable}
\usepackage{makecell}
\usepackage{booktabs}
\usepackage[hyphens]{url}
\usepackage[colorlinks=true, allcolors=blue, breaklinks=true]{hyperref}
\usepackage{multirow}

\usepackage{color,soul}

\usepackage{scalerel}
\usepackage{tikz}
\usetikzlibrary{svg.path}

\definecolor{orcidlogocol}{HTML}{A6CE39}
\tikzset{
  orcidlogo/.pic={
    \fill[orcidlogocol] svg{M256,128c0,70.7-57.3,128-128,128C57.3,256,0,198.7,0,128C0,57.3,57.3,0,128,0C198.7,0,256,57.3,256,128z};
    \fill[white] svg{M86.3,186.2H70.9V79.1h15.4v48.4V186.2z}
                 svg{M108.9,79.1h41.6c39.6,0,57,28.3,57,53.6c0,27.5-21.5,53.6-56.8,53.6h-41.8V79.1z M124.3,172.4h24.5c34.9,0,42.9-26.5,42.9-39.7c0-21.5-13.7-39.7-43.7-39.7h-23.7V172.4z}
                 svg{M88.7,56.8c0,5.5-4.5,10.1-10.1,10.1c-5.6,0-10.1-4.6-10.1-10.1c0-5.6,4.5-10.1,10.1-10.1C84.2,46.7,88.7,51.3,88.7,56.8z};
  }
}

\newcommand\orcidicon[1]{\href{https://orcid.org/#1}{\mbox{\scalerel*{
\begin{tikzpicture}[yscale=-1,transform shape]
\pic{orcidlogo};
\end{tikzpicture}
}{|}}}}

\newcommand{\verspt}{\\[0.1em]}
\newcommand{\verspb}{\\[0.2em]}
\usepackage{enumitem}
\setlist[itemize]{noitemsep, topsep=0pt}
\usepackage{hyperref} %<--- Load after everything else

\title{6GENABLERS-DLT: DLT-based Marketplace for Decentralized 
Trading of 6G Telco resources}
\author{
Adriana Fernández-Fernández \orcidicon{0000-0003-1616-5582}, Researcher at i2CAT \textit{adriana.fernandez@i2cat.net} \and 
Angel Martin \orcidicon{0000-0002-1213-6787}, Researcher at Vicomtech \textit{amartin@vicomtech.org} \and
Guillermo Gomez, Researcher at ATOS \textit{guillermo.gomezchavez@eviden.com}
}

\usepackage{xcolor}
\usepackage{svg}

\usepackage{fancyhdr}% http://ctan.org/pkg/fancyhdr
\usepackage{titlesec}
\titleformat{\paragraph}
{\normalfont\normalsize\bfseries}{\theparagraph}{1em}{}
\titlespacing*{\paragraph}
{0pt}{3.25ex plus 1ex minus .2ex}{1.5ex plus .2ex}

\begin{document}
\maketitle
\begin{abstract}
The 6GENABLERS-DLT project addresses critical challenges in fostering multi-party collaboration within dynamic 6G environments. As operators and service providers increasingly depend on third-party resources to meet their contractual and operational needs, the project introduces an innovative, Distributed Ledger Technology (DLT)-anchored Marketplace designed to streamline decentralized telco resource trading. This 6GENABLERS Marketplace serves as a collaborative platform where operators, resource providers, and service providers can seamlessly discover, advertise, and trade telco assets within a transparent, secure, and efficient permissioned environment. Distinguished from public DLT/Blockchain solutions, the Marketplace’s permissioned nature ensures robust governance, privacy, and control, making it particularly suited to enterprise and consortium-based use cases in the Information and Communication Technology (ICT) sector. The adoption of a decentralized architecture eliminates reliance on a central operator, thereby mitigating risks associated with single points of failure and enhancing system trustworthiness, resilience, and fault tolerance. The Marketplace encompasses a wide range of resources integral to 6G networks, including virtualized mobile core components, Radio Access Network (RAN) assets, edge and cloud infrastructure, and vertical applications tailored to specific industry needs. This diversity enables stakeholders to dynamically access and scale resources, fostering operational efficiency and innovation across 6G ecosystems. Through the 6GENABLERS-DLT project, the vision of a collaborative, resource-rich 6G environment becomes a reality, laying the foundation for a next-generation telco ecosystem where decentralization empowers stakeholders to meet the demands of an interconnected, flexible, and scalable future.
\end{abstract}

\section{6GENABLERS-DLT vision}
\label{sec:vision}
% Brief overview of the 6GENABLERS-DLT project goals, scope, and key outcomes.
% Highlight the relevance of a DLT-anchored distributed telco marketplace for 6G networks.
% Summary of the project's main achievements, including prototypes, validation results, and practical use cases.
% Motivation for the project: challenges in multi-party resource sharing and collaboration in 6G.
% Importance of decentralization, DLTs, and smart marketplaces for the future of telecommunications.
% Key objectives and scope of the 6GENABLERS-DLT project.

The 6GENABLERS-DLT project is a groundbreaking initiative aimed at addressing the challenges of multi-party collaboration in the emerging 6G ecosystem. As 6G networks evolve to accommodate increasingly complex and dynamic environments, operators and service providers must often rely on third-party resources to meet diverse operational demands. The 6GENABLERS-DLT project introduces a decentralized, Distributed Ledger Technology (DLT)-anchored Smart Marketplace that empowers stakeholders to discover, advertise, and trade telco resources in a transparent, secure, and efficient manner.

The 6GENABLERS-DLT Marketplace stands out for its decentralization and permissioned architecture, ensuring robust governance, privacy, and control. Unlike public Blockchain solutions, this tailored approach enhances security and transparency while meeting the unique demands of enterprise and consortium-driven applications.

By offering a diverse range of resources—including virtualised mobile core components, Radio Access Network (RAN) assets, edge and cloud infrastructures, and industry-specific applications—the Marketplace promotes collaboration, flexibility, and scalability across the 6G landscape. Operators and resource/service providers can seamlessly access these resources to optimize operations and deliver enhanced services to end consumers.

Through this transformative project, the 6GENABLERS-DLT initiative lays the groundwork for a highly interconnected, resource-rich 6G ecosystem, fostering technological innovation and operational excellence for the next generation of communication networks.

The motivation behind the 6GENABLERS-DLT project stems from the challenges inherent in dynamic 6G environments, where operators and service providers often depend on third-party resources to meet diverse operational needs. Traditional centralized models for resource sharing are insufficient, plagued by inefficiencies, lack of transparency, and a heightened risk of single points of failure.  

To address these challenges, 6GENABLERS-DLT envisioned and developed a DLT-anchored Smart Marketplace, tailored to facilitate seamless, secure, and efficient resource trading. This decentralized platform mitigates trust issues, ensures fault tolerance, and fosters a collaborative ecosystem where telco resources—including virtualised core components, RAN assets, and edge/cloud infrastructure—can be shared effectively.  

The project’s key goals include:  
\begin{enumerate}
    \item Creating a robust, permissioned decentralized Marketplace for telco resource discovery and trading.
    \item Developing advanced tools for DLT integration, identity management, and Service Level Agreement (SLA) monitoring.
    \item Demonstrating the feasibility and utility of the Marketplace through prototypes, validations, and practical use cases.
    \item Driving decentralisation and reducing reliance on centralized intermediaries.
\end{enumerate}

The project has delivered significant advancements: 
\begin{itemize}
    \item Prototypes: Functional demonstrations of the Marketplace, showcasing its ability to advertise, trade, and monitor resources.
    \item Validation Results: Rigorous testing of the DLT-based architecture, confirming its efficiency, scalability, and fault tolerance.
    \item Use Cases: Practical scenarios highlighting the Marketplace’s role in multi-party collaboration, such as dynamic resource allocation and SLA-based trading in complex 6G environments.  
\end{itemize}

The 6G ecosystem demands unprecedented levels of flexibility, scalability, and trust. The Marketplace of 6GENABLERS-DLT exemplifies how DLTs and Smart Marketplaces can address these needs by: ensuring trustless operations through tamper-proof records; enabling privacy-conscious and permissioned environments for enterprise-grade applications; supporting resilient decentralized architectures that eliminate single points of failure.  

By developing a DLT-anchored distributed Marketplace, 6GENABLERS-DLT has laid the foundation for the next generation of collaborative and decentralized telecommunications infrastructure, paving the way for innovative business models and enhanced operational efficiencies in the 6G era.

This project has been made possible thanks to funding provided by the Spanish Ministry of Economic Affairs and Digital Transformation and the European Union – NextGenerationEU, as part of the Recovery, Transformation, and Resilience Plan (PRTR). The funding was secured under the UNICO I+D 5G 2021 Call. This financial support has enabled the \href{https://i2cat.net/}{i2CAT} Foundation to spearhead the development of this innovative Marketplace, leveraging cutting-edge technologies and expertise.

As part of the initiative, two tenders were awarded to industry leaders to ensure the success of the project:

\begin{itemize}
    \item \href{https://www.vicomtech.org/en}{Vicomtech} was tasked with developing the core DLT system, trading Application Programming Interfaces (APIs), identity management, data bus, network orchestration brokerage, SLA monitoring, prediction, and alarm systems.
    \item \href{https://eviden.com/}{ATOS} was responsible for creating tools for smart contract generation and building intuitive console interfaces for human operators managing the distributed Marketplace nodes.
\end{itemize}

This paper is structured as follows. Section \ref{sec:usecases} overviews identified use cases and derived requirements for the marketplace. Section \ref{sec:architecture} describes the final architecture design. Section \ref{sec:prototypes} details the developed prototypes and the specialised features provided. Section \ref{sec:validation} reports the tests done around the target use cases. Section \ref{sec:lessons} lists the lessons learned relevant for 6G stakeholders. Section \ref{sec:impact} sums up the dissemination activities. Finally, Section \ref{sec:conclusions} provides a few highlights.

\section{Use Cases and Requirements}
\label{sec:usecases}
% Overview of deliverable E2.1: identified use cases and derived requirements for the marketplace.
% Explanation of the operational challenges addressed, including SLA assurance, trading models, and resource provisioning.
% Example use cases:
% VNF Trading
% Edge/Cloud Resource Trading
% Network Slice Trading
% Network Service Trading

6GENABLERS-DLT project identifies the use cases and derives the requirements for the Distributed Telco Marketplace, emphasizing its role as a transformative solution for multi-party collaboration in 6G networks. The project elaborates on the operational challenges within the scope of SLA assurance, trading models, and resource provisioning, offering a comprehensive analysis of how these aspects are addressed through the proposed marketplace.

The 6GENABLERS Marketplace addresses critical challenges in 6G ecosystems, including:  
\begin{itemize}
    \item SLA Assurance: Guaranteeing adherence to SLA terms, with mechanisms for monitoring, predicting, and resolving violations.  
    \item Trading Models: Enabling seamless and decentralized transactions through Smart Contracts, DLTs, and a permissioned environment that fosters trust and transparency.  
    \item Resource Provisioning: Facilitating dynamic allocation and provisioning of infrastructure components, such as Virtual Network Functions (VNFs), edge/cloud resources, and network slices, tailored to the needs of service providers and operators.  
\end{itemize}

\begin{figure*}
\centering
  \includegraphics[width=0.99\textwidth]{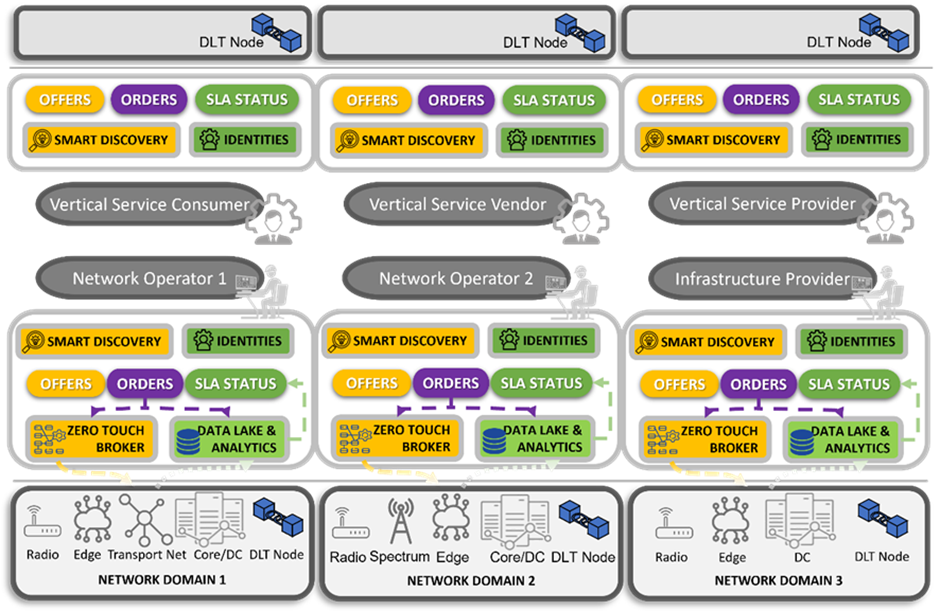}
\caption{Actors in the 6GENABLERS Marketplace.}
\label{fig:actors} % Give a unique label
\end{figure*}

The 6GENABLERS Marketplace involves a variety of stakeholders, each playing essential roles, depicted in Figure \ref{fig:actors}: 
\begin{enumerate}
    \item Vertical Service Providers (VSPs): Central to delivering holo-comm services, VSPs utilize Marketplace resources such as edge computing, network connectivity, and VNFs to enable real-time sessions. They engage with Network Operators, Infrastructure Providers, and Vertical Service Vendors (VSVs) to acquire and integrate these assets.  
    \item Network Operators: Providers of network infrastructure, e.g., 5G, Beyond 5G (B5G), 6G, to ensure high-quality, low-latency connectivity. They participate in the Marketplace by offering network resources to VSPs.  
    \item Infrastructure Providers: Supply additional resources like edge computing, storage, and connectivity required for real-time holographic communications, expanding beyond Network Operators' offerings.  
    \item Vertical Service Vendors (VSVs): Develop and trade VNFs tailored for holo-comm environments, engaging with VSPs to enable innovative services.  
    \item Vertical Service Consumers (VSCs): End users who benefit from the holo-comm services provided by VSPs, driving demand in Business-to-Customer (B2C) transactions.  
    \item Governance Board: Oversees Marketplace policies, ensures integrity, and manages participant onboarding and dispute resolution.  
\end{enumerate}

This collaborative framework supports efficient resource trading, enabling scalable and sustainable holographic communication services within the 6GENABLERS Marketplace.

6GENABLERS-DLT outlines several example use cases to illustrate the practical application of the Distributed Telco Marketplace, including:  
\begin{itemize}
    \item VNF Trading: Demonstrates the trading of Virtual Network Functions (VNFs) such as edge compute components or core network functions, including the steps for stakeholder onboarding, offer creation, offer discovery, order creation, and order deployment.  
    \item Edge/Cloud Resource Trading and Provisioning: Details the acquisition and deployment of edge and cloud resources to host application orchestrators and other services, highlighting the interplay between SLA monitoring and resource provisioning.  
    \item Network Slice Trading and Provisioning: Focuses on trading network slices encompassing RAN, Core, and Edge capabilities, essential for XR services requiring specific QoS guarantees. This use case includes SLA violation detection and notification mechanisms.  
    \item Network Service Trading and Provisioning: Illustrates the process of acquiring complete network services, including their deployment and SLA enforcement.  
\end{itemize}

From these use cases, a comprehensive set of requirements for the marketplace's functionalities has been extracted. These include:
\begin{itemize}
    \item Ensuring secure onboarding of stakeholders and management of distributed identities.  
    \item Enabling efficient and automated discovery, trading, and provisioning of resources.  
    \item Guaranteeing SLA compliance through real-time monitoring and predictive analytics.  
\end{itemize}

The 6GENABLERS Marketplace is defined by a set of functional and non-functional requirements aimed at ensuring robust, scalable, and efficient operation. These requirements address the core functionalities of the Marketplace, including trading mechanisms, identity management, and SLA assurance, as well as system-wide attributes like security, scalability, and interoperability.  

The Functional requirements focus on the essential capabilities the Marketplace must deliver. For example, trading-related requirements include mechanisms for offer creation, discovery, and order management, ensuring seamless interactions between stakeholders. Identity management requirements emphasize secure onboarding, role-based access control, and support for distributed identities to link resources, such as VNFs, network slices, and cloud resources, to their respective stakeholders. SLA monitoring requirements ensure real-time evaluation of agreed-upon performance metrics, with features like proactive alerts for SLA violations and forecasting capabilities.  

The Non-Functional requirements focus on overarching system attributes that ensure the Marketplace's usability and performance. For instance, scalability requirements emphasize the ability to handle increasing numbers of offers, orders, and participants without degrading performance. Security requirements underline the need for end-to-end encryption, secure authentication, and tamper-proof records via DLTs. Interoperability requirements ensure seamless integration with existing orchestration frameworks like i2Slicer \cite{i2slicer} and VINO, as well as compatibility with standards such as TM Forum APIs \cite{tmf}.  

Some Key Performance Indicators (KPIs) are built on top of these requirements, defined to measure the success of the platform and its ability to meet operational and functional goals. These KPIs include:  
\begin{itemize}
    \item Latency in SLA Monitoring and Alerts: Ensures SLA violations are detected and reported within predefined timeframes, enabling timely corrective actions.
    \item Throughput of Trading Operations: Measures the number of transactions, such as offers and orders, that the system can process per second.
    \item Success Rate of Offer Discovery: Evaluates the system's ability to match consumer intents with appropriate resources, ensuring a high percentage of successful matches.
    \item Scalability Benchmarks: Assesses the system’s performance under high-load scenarios, ensuring resource utilization remains efficient and responsive.  
\end{itemize}

For example, a representative functional requirement might involve linking offers and orders to SLA terms through distributed identities, while a corresponding KPI measures the accuracy and timeliness of SLA violations being flagged. Similarly, a non-functional requirement for seamless integration with orchestration systems is reflected in KPIs evaluating deployment success rates for network slices or VNFs.  

Together, these requirements and KPIs provide a comprehensive framework for validating the 6GENABLERS Marketplace’s ability to facilitate decentralized, SLA-driven trading of 6G infrastructure resources.

By addressing these challenges and meeting the outlined requirements, the 6GENABLERS Marketplace aims to redefine resource sharing and trading paradigms in 6G networks, creating a robust, decentralized, and efficient ecosystem.

\section{Architecture of the 6GENABLERS Marketplace}
\label{sec:architecture}
% Summary of deliverable E2.3: final architecture design.
% Key components and layers:
% Business Layer: Smart Contracts, offer/order management, identity management.
% Monitoring and Control Layer: SLA tracking, resource orchestration, and service instantiation.
% Integration with Network Infrastructure.
% High-level diagram of the architecture for clarity.

6GENABLERS-DLT has produced an architecture of the 6GENABLERS Marketplace, which forms the backbone for efficient and secure resource trading in 6G networks. The Marketplace is designed as a decentralized platform built on DLT, enabling stakeholders to trade various 6G infrastructure resources.

The architecture consists of two primary layers:

\begin{enumerate}
    \item Business Layer: This layer manages core marketplace operations, including Smart Contracts, offer/order management, and Identity Management. It facilitates secure transactions and the creation of offers and orders within the marketplace, ensuring that all actions are compliant with the established business logic and protocols.
    \item Monitoring and Control Layer: This layer handles SLA tracking, resource orchestration, and service instantiation. It ensures the operational integrity of the system, tracking performance metrics and ensuring that services meet predefined quality standards. This layer is essential for maintaining trust and compliance in the decentralized marketplace environment.
\end{enumerate}

The 6GENABLERS Marketplace is designed to integrate seamlessly with the underlying network infrastructure, supporting both edge and cloud resources, along with network slices. This allows the Marketplace to facilitate the provisioning of network services, ensuring that resources are properly allocated and monitored to meet the demands of real-time, mission-critical applications like holographic communications.

\begin{figure*}
\centering
  \includegraphics[width=0.6\textwidth]{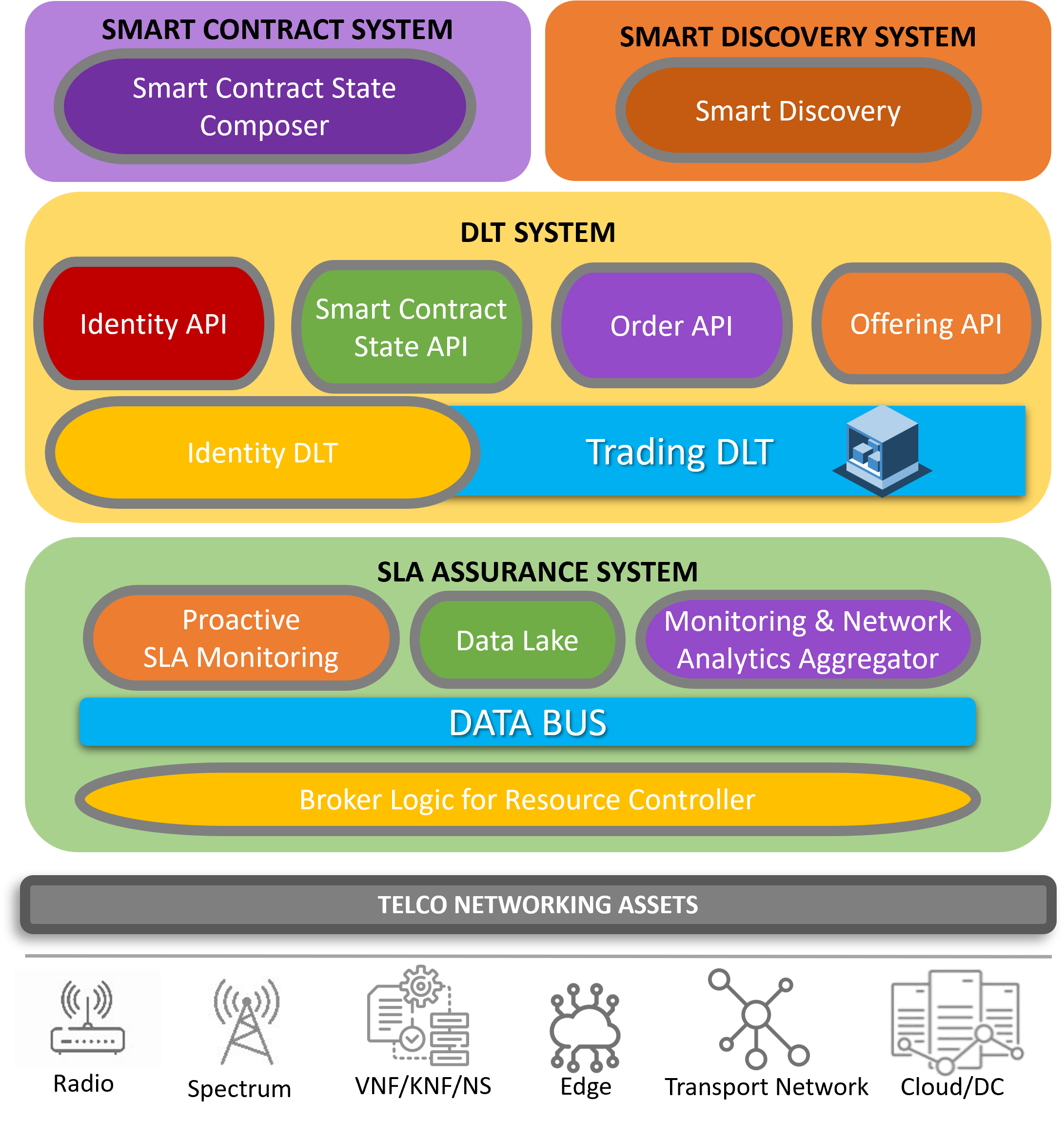}
\caption{6GENABLERS Marketplace high-level architecture.}
\label{fig:arch} % Give a unique label
\end{figure*}

The Figure \ref{fig:arch} in the document presents the high-level architecture of the 6GENABLERS Marketplace, depicting the two main layers: the Business Layer and the Monitoring and Control Layer, along with their integration with the Network Infrastructure. This diagram provides a comprehensive view of how the different components interact within the Marketplace ecosystem. 

The components present in the architecture are:
\begin{enumerate}
    \item Trading DLT \& Identity DLT: These components provide the decentralized ledger infrastructure for managing trading transactions and identities. The Trading DLT handles the marketplace transactions, while the Identity DLT manages decentralized identifiers for secure and permissioned access. The identity API manages identity-related actions, including issuing, validating, and revoking Decentralized Identifiers (DIDs) for users and services in the marketplace. It ensures secure and privacy-preserving authentication.
    \item Offer, Order and SC State API: These APIs handle core marketplace operations. The Offering API manages resource listings, the Order API facilitates service requests, and the SC State API tracks smart contract states related to resource deployments.
    \item Smart Contract State Composer: This component is responsible for composing and managing smart contract states, integrating legal terms with technical terms for automation in the marketplace.
    \item Smart Discovery: This application uses machine learning algorithms to enable smart and efficient discovery of available offers based on user intent, helping match consumers with suitable resources.
    \item Data Bus: A messaging ensures communication between components, providing an asynchronous channel for sharing updates and events across the marketplace's different nodes.
    \item Broker Logic for Resource Controller: controls resource orchestration, translating marketplace orders into actionable commands for resource deployment using the appropriate network and computing systems.
    \item Monitoring \& Network Analytics Aggregator: aggregates monitoring data from the network to track system performance and resource utilization, essential for SLA enforcement.
    \item Data Lake: A central repository for storing large amount of data. It ensures that performance metrics are available for analysis and auditing .
    \item Proactive SLA Monitoring: This system continuously evaluates SLA and predicts potential SLA violations, enabling timely interventions to avoid breaches.
\end{enumerate}

These components form the backbone of the 6GENABLERS Marketplace, working together efficient, secure, and transparent trading of resources within a decentralized 6G ecosystem. Each module plays a vital role in ensuring the system's functionality, scalability, and compliance with defined SLAs.

By employing DLT for secure, transparent transactions and SLA assurance mechanisms, the architecture supports a robust, scalable environment that meets the demands of 6G environments. The inclusion of Smart Contracts, Identity Management, and Proactive SLA Monitoring ensures the Marketplace can handle complex, multi-party resource sharing in a decentralized manner.

\section{Prototypes and Technical Innovations}
\label{sec:prototypes}

\subsection{DLT Testbed Implementation}
\label{sec:dlt}
% Summary of deliverable E3.3: technical details of the implemented DLT testbed.
% Selection process for DLT technologies (Hyperledger Fabric, Quorum, Corda).
% Evaluation criteria (e.g., scalability, performance, consensus algorithms).

6GENABLERS-DLT builds a DLT testbed for the 6GENABLERS Marketplace. This testbed represents a foundational component in enabling a decentralized, transparent, and efficient trading system for 6G networks. Along different project milestones, 6GENABLERS-DLT has elaborated on the technology selection, development processes, and evaluation results for the DLT testbed.

The selection of DLT technologies—Hyperledger Fabric, Quorum, and Corda—was based on their suitability for key marketplace requirements. The evaluation process considered criteria such as scalability, performance, consensus algorithms, privacy capabilities, and support for permissioned environments. Hyperledger Fabric was ultimately chosen as the primary technology due to its modular architecture, strong privacy controls, and robust support for enterprise-grade distributed applications.

The testbed's architecture was designed to facilitate decentralized trading while ensuring consistency and scalability across multiple network nodes. Key features include:
\begin{enumerate}
    \item Hyperledger Fabric Deployment: The testbed leverages Kubernetes for deploying Hyperledger Fabric, enabling the creation of a distributed network of peer nodes. Each node hosts key components such as orderers, chaincodes, and CouchDB for distributed data storage.
    \item Chaincode Development: Custom chaincodes were implemented to support specific marketplace functions, including offer publication, order management, and Smart Contract (SC) state updates. These chaincodes facilitate data integrity and transactional transparency.
    \item Data Model Alignment: The APIs and chaincodes follow a JSON-based data model compatible with TM Forum standards, ensuring seamless integration with external systems and maintaining interoperability across the marketplace.
    \item Automation and Onboarding: Automated installation scripts and onboarding procedures were developed to streamline the deployment of DLT nodes and ensure secure integration of new marketplace participants.
\end{enumerate}

\begin{figure*}
\centering
  \includegraphics[width=0.7\textwidth]{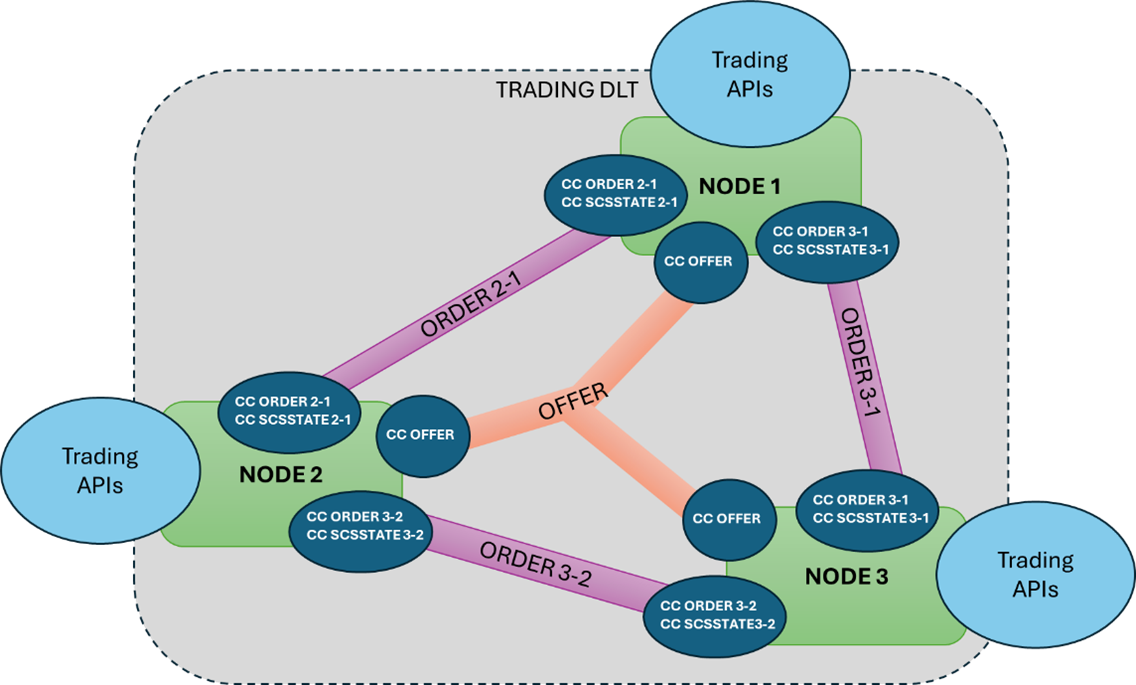}
\caption{Trading DLT high-level architecture.}
\label{fig:dltarch} % Give a unique label
\end{figure*}

All these features compose the distributed infrastructure using channels to publicly share offers and privately make order transactions and report progress status and SLA alarms. This structure is depicted in Figure \ref{fig:dltarch}.

The implemented testbed underwent rigorous functional and integration tests to validate its capabilities. These tests evaluated the system's ability to process transactions, enforce privacy policies, and synchronize data across the distributed network. Additionally, performance benchmarks assessed metrics such as transaction latency, throughput, and resource utilization, confirming the testbed's readiness for real-world applications.

The onboarding and maintenance procedures for DLT nodes in the 6GENABLERS Marketplace are critical for ensuring secure, consistent, and efficient integration of new nodes into the distributed ledger network. These procedures are structured into two scenarios based on the role of the node being onboarded: Admin Nodes and Non-Admin Nodes.

Admin nodes are pivotal in managing the governance and operational integrity of the DLT network. The onboarding process follows a stringent procedure to guarantee the authenticity and security of these nodes. While, non-admin nodes represent standard participants in the DLT network, such as trading or monitoring nodes. Their onboarding process leverages the same repository-based mechanism but with reduced privileges and responsibilities.

A candidate admin node submits a repository commit request containing its public certificate and metadata required for integration. Existing admin users of the repository review the commit request. The admin users either approve or reject the request. Approved nodes are added to the repository with their associated certificates. Once the commit is accepted, all existing nodes in the network automatically update their configurations using the approved certificates. This ensures seamless connectivity between the new admin node and the rest of the network.

The key highlights of the onboarding mechanism include:
\begin{itemize}
    \item Repository-Centric Workflow: The onboarding process is centralized through a repository, ensuring transparency and consistency in node integration.
    \item Automated Certificate Distribution: Upon acceptance of a commit request, the certificates of new nodes are propagated across the network, enabling immediate connectivity with all existing nodes.
    \item Scalable and Secure Integration: This approach supports the seamless addition of nodes while maintaining high security and trustworthiness in the DLT network.
\end{itemize}

\subsection{Smart Contracts for Resource Negotiation}
\label{sec:smartcontracts}
% Overview of deliverable E4.3: implementation of negotiation processes using Smart Contracts.
% Role of automation and AI/ML in enhancing negotiation and matching offers to user intents.

6GENABLERS-DLT provides an in-depth exploration of the implementation of negotiation processes leveraging Smart Contracts (SCs) within the 6GENABLERS Marketplace. This work focuses on the integration of SCs as a mechanism for facilitating secure, transparent, and efficient resource and service trading in distributed 6G networks. This section outlines the SC system's architecture, key components, data models, and practical use cases for resource negotiation.

First, it is important to know the two different business scenarios which are considered:
\begin{itemize}
    \item Business-to-Business (B2B) Scenario: For example, a Vertical Service Vendor (VSV) might use the Marketplace to sell VNFs with fixed recurring costs. The SC Dev Toolkit prepares associated documents like License Models and Legal Prose Templates, integrating them into the AIOs.
    \item Business-to-Customer (B2C) Scenario: A Vertical Service Provider (VSP) may offer access to VNFs, where consumers acquire resources and services through smart contracts. In these cases, the Distributed Identifiers (DIDs) act as references for enhanced traceability and security.
\end{itemize}

To generate and handle the data models required in these two scenarios, the SC system's architecture includes, two systems. The Smart Contract Developer Toolkit (SC Dev Toolkit) and the Smart Contract State Composer (SCSC).

\begin{figure*}
\centering
  \includegraphics[width=0.8\textwidth]{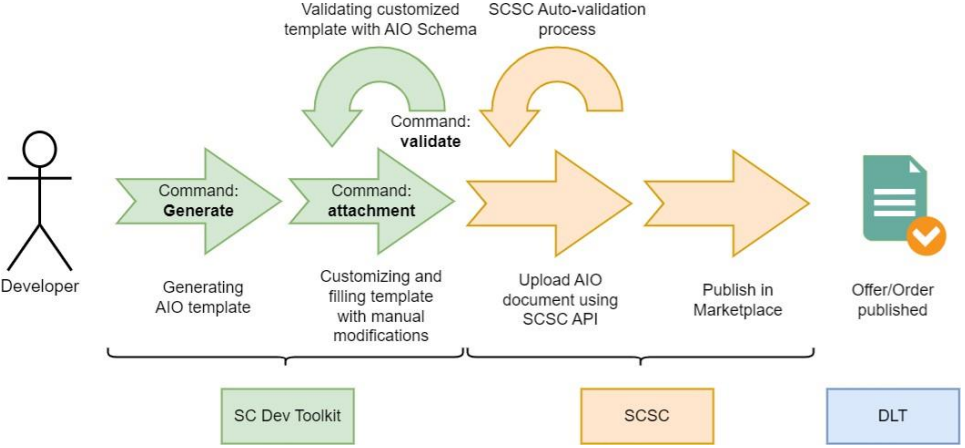}
\caption{Model creation process using SC Dev Toolkit.}
\label{fig:scsarch} % Give a unique label
\end{figure*}

The SC Dev Toolkit is an external Command Line Interface (CLI) tool that supports the creation of All-In-One (AIO) data models in JavaScript Object Notation (JSON) format. These AIOs encapsulate the offering, ordering, and Smart Contract states. Through an iterative process (illustrated in Figure \ref{fig:scsarch}), developers begin with baseline templates and adapt them to reflect specific requirements. The AIOs are validated for integration into the SC State Composer (SCSC), ensuring consistency and interoperability across the Marketplace.

The SCSC is the core component of the SC system, deployed alongside the other Marketplace components. It facilitates seamless interaction with the DLT and provides an entry point for stakeholders to engage in trading activities. The SCSC ensures that stakeholders can create, manage, and monitor SC states while integrating features like legal binding and lifecycle management. 

Key functionalities of the SCSC include:
\begin{itemize}
    \item Creation of SC States: Establishing a structured representation of orders during instantiation.
    \item Legal Prose Generation: Rendering transaction details into a Portable Document Format (PDF) format that provides human-readable legal documentation.
    \item Lifecycle Management (LCM): Managing the complete lifecycle of offers and orders, including updates triggered by SLA monitoring or availability changes.
\end{itemize}

The SC states considered are messages structured as events are the following:
\begin{itemize}
    \item SLA Violation Prediction: fired as an "Alert" message by the Proactive SLA Monitoring system when a potential SLA Violation has been predicted by the platform.
    \item SLA Violation: fired as an "Alarm" message by the Proactive SLA Monitoring system when any of the conditions or rules of the SLA have been breached.
    \item Offering Violation: fired as an "offering\_violation" message by the xNF Event Exposure module when any of the conditions (e.g., pricing payment periods, storage or CPU values, etc) of the Offering have been breached.
    \item Availability: fired as an "availability\_response" message by the Broker Logic for Resource Controller according to the monitored availability of the resources of an offer.
    \item Deployment Status: fired as a "deployment\_status" message by the Broker Logic for Resource Controller when reporting the status of an order being deployed.
\end{itemize}

The 6GENABLERS Marketplace supports trading scenarios across diverse resource types, such as VNFs, Network Services (NSs), Edge/Cloud resources, RAN services, and network slices. The flexibility of TM Forum models ensures adaptability to the specific characteristics of each resource type, enabling the representation of technical parameters like cores, RAM, frequencies, or operating bands.

By leveraging DLT, the SC system ensures decentralized management of transaction data, enabling trust and transparency among stakeholders. The AIOs use modified TM Forum schemas to include only relevant DIDs, ensuring alignment with the permissioned nature of the Marketplace. This adaptation streamlines negotiation processes while maintaining compatibility with TM Forum standards. 

The Smart Contracts integrate seamlessly with the Marketplace's operational components, driving automation, trust, and efficiency in resource trading. Through these mechanisms, the 6GENABLERS Marketplace establishes a robust foundation for multi-stakeholder collaboration in future 6G networks.

\subsection{Smart Discovery for Intent-based selection}
\label{sec:smartdiscovery}

The Smart Discovery system in the 6GENABLERS Marketplace consists of two core components: the Clustering and Classification module and the Intent-based Discovery module, both designed to facilitate intelligent resource and service discovery. These components aim to align user intents with available offerings in a seamless, efficient, and intuitive manner.

The Clustering and Classification module organizes a wide variety of offers by identifying patterns within them and grouping similar offers into clusters. This module leverages advanced machine learning algorithms, including unsupervised methods like Factor Analysis of Mixed Data (FAMD) and K-means, for offline clustering. It also uses supervised learning to classify new offers dynamically at runtime, ensuring each offer is mapped to the appropriate cluster for future retrieval. This clustering improves the efficiency of the discovery process and supports rapid classification for high scalability in the 6GENABLERS Marketplace.

Some key features include:
\begin{itemize}
    \item Pattern Discovery: Recognizing similarities across resource offers.
    \item New Offer Classification: Assigning new offers to relevant clusters for efficient searching.
    \item Cluster Resolution: Identifying and retrieving the most relevant clusters based on user intents, enabling tailored recommendations.
\end{itemize}

\begin{figure*}
\centering
  \includegraphics[width=0.6\textwidth]{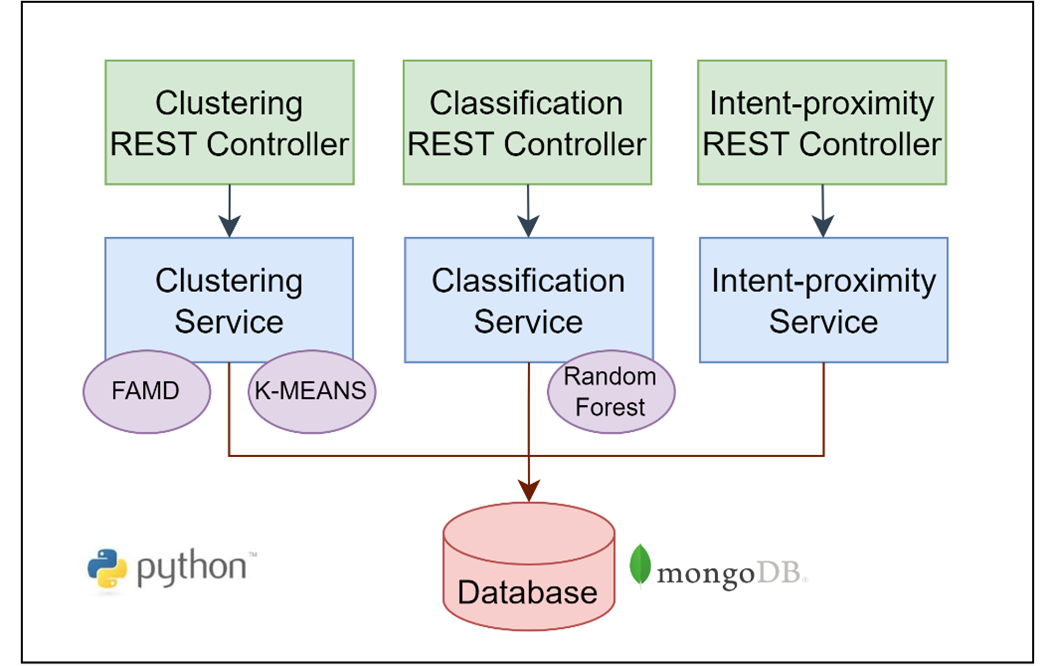}
\caption{Software architecture of the Clustering and Classification module.}
\label{fig:sdclustarch} % Give a unique label
\end{figure*}

The architecture of the Clustering and Classification module is depicted in Figure \ref{fig:sdclustarch}.

The Intent-based Discovery module translates user input, whether in natural or semi-natural language, into actionable queries. By integrating with large language models (LLMs), it refines the understanding and processing of user intents. This module interacts with the Clustering and Classification module to retrieve offers aligned with user preferences, offering functionalities such as:
\begin{itemize}
    \item Translating intents into queryable structures.
    \item Identifying relevant clusters based on user input.
    \item Retrieving offers from the marketplace for further filtering and sorting.
\end{itemize}

\begin{figure*}
\centering
  \includegraphics[width=0.5\textwidth]{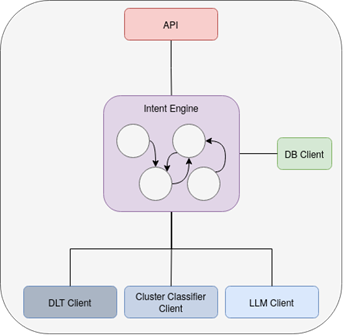}
\caption{Software architecture of the Intent-based Discovery module.}
\label{fig:sdibarch} % Give a unique label
\end{figure*}

The architecture of the Intent-based Discovery, depicted in Figure \ref{fig:sdibarch}, demonstrates its modular design, which allows seamless integration with other Marketplace components, such as the Offering API and the Trading DLT.

The key internal components are:
\begin{itemize}
    \item API Gateway: Handles all incoming requests, processing natural or semi-natural language intents.
    \item Database Client: Manages storage and retrieval of intents and offers.
    \item Intent Engine: Serves as the controller, coordinating operations among components.
    \item Southbound Clients: Connect to external APIs for tasks such as clustering, classification, and data retrieval from the DLT.
\end{itemize}

This workflow ensures the discovery process remains efficient and user-centric, enhancing the overall functionality of the 6GENABLERS Marketplace. It provides stakeholders with a robust, intelligent system to locate and acquire resources that meet their specific requirements.

\subsection{Operational Components for SLA Assurance}
\label{sec:sla}
% Highlights from E5.3: SLA monitoring, prediction, and enforcement mechanisms.
% Key functionalities: proactive alerts, Smart Contract integration, and analytics.

The 6GENABLERS Marketplace SLA Assurance System is a cornerstone of the platform, designed to monitor, predict, and enforce SLA compliance across distributed telco infrastructures. Building upon the deliverable E5.3, this section details the key functionalities and implementation of the SLA Assurance operational components, which ensure proactive, transparent, and reliable management of service quality agreements.

The SLA Assurance system supports local monitoring, predictive analytics, and real-time SLA condition enforcement at each node of the distributed Marketplace. By integrating advanced monitoring tools with DLT capabilities, this system enables:
\begin{itemize}
    \item Proactive SLA Alerts: Utilization of predictive algorithms to forecast potential SLA breaches before they occur.
    \item Integration with Smart Contracts: Enforcement of SLA conditions via smart contract-triggered workflows, ensuring compliance with predefined quality metrics.
    \item Data-Driven Analytics: A robust mechanism for evaluating SLA adherence, identifying deviations, and enabling informed decision-making.
\end{itemize}

\begin{figure*}
\centering
  \includegraphics[width=0.8\textwidth]{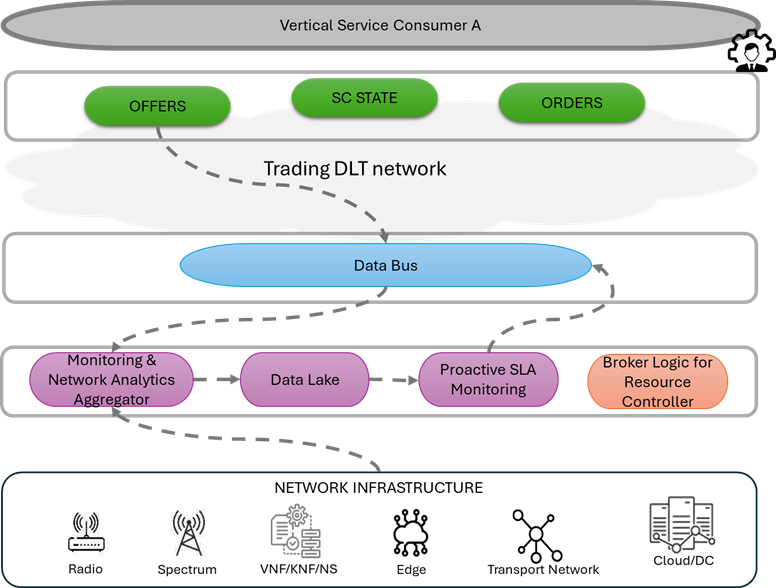}
\caption{Proactive SLA Monitoring overview architecture.}
\label{fig:slaarch} % Give a unique label
\end{figure*}

The workflow for proactive SLA Monitoring is depicted in Figure \ref{fig:slaarch}.

The key functionalities implemented are:
\begin{itemize}
    \item Proactive Alerts: The system employs predictive analytics to evaluate live metrics against SLA conditions, issuing alerts when breaches are anticipated, thus enabling timely remedial actions.
    \item Smart Contract Integration: SLA rules and metrics are encoded within TM Forum-compliant smart contracts. The system continuously tracks compliance and logs violations directly into the DLT for auditability.
    \item Advanced Analytics and Visualization: Metrics from SLA monitoring are stored in the Data Lake, allowing comprehensive analytics through tools like Prometheus and Grafana. This facilitates intuitive visualization of SLA compliance trends.
\end{itemize}

By incorporating state-of-the-art SLA monitoring and predictive capabilities, the operational components of the SLA Assurance system significantly enhance the reliability and resilience of resource trading within the 6GENABLERS Marketplace. These innovations position the Marketplace as a leader in enabling transparent and efficient multi-party collaborations in 6G environments.

\section{Validation and Performance Analysis}
\label{sec:validation}
% Insights from deliverable E6.2: validation setup, test scenarios, and performance results.
% Coverage of KPIs and KVIs for:
% SLA Monitoring and Notification.
% Scalability of the marketplace.
% Resilience of the DLT-based architecture.
% Outcomes from test cases in the identified use cases.

This section provides insights from the validation setup, test scenarios, and performance results for the 6GENABLERS Marketplace. The focus lies on validating SLA Monitoring and Notification, Marketplace Scalability, and the Resilience of the DLT-based architecture. The analysis includes outcomes from key test cases executed across the identified use cases: VNF Trading, Edge/Cloud Resource Trading and Provisioning, Network Slice Trading and Provisioning, and Network Service Trading and Provisioning.

\begin{figure*}
\centering
  \includegraphics[width=0.9\textwidth]{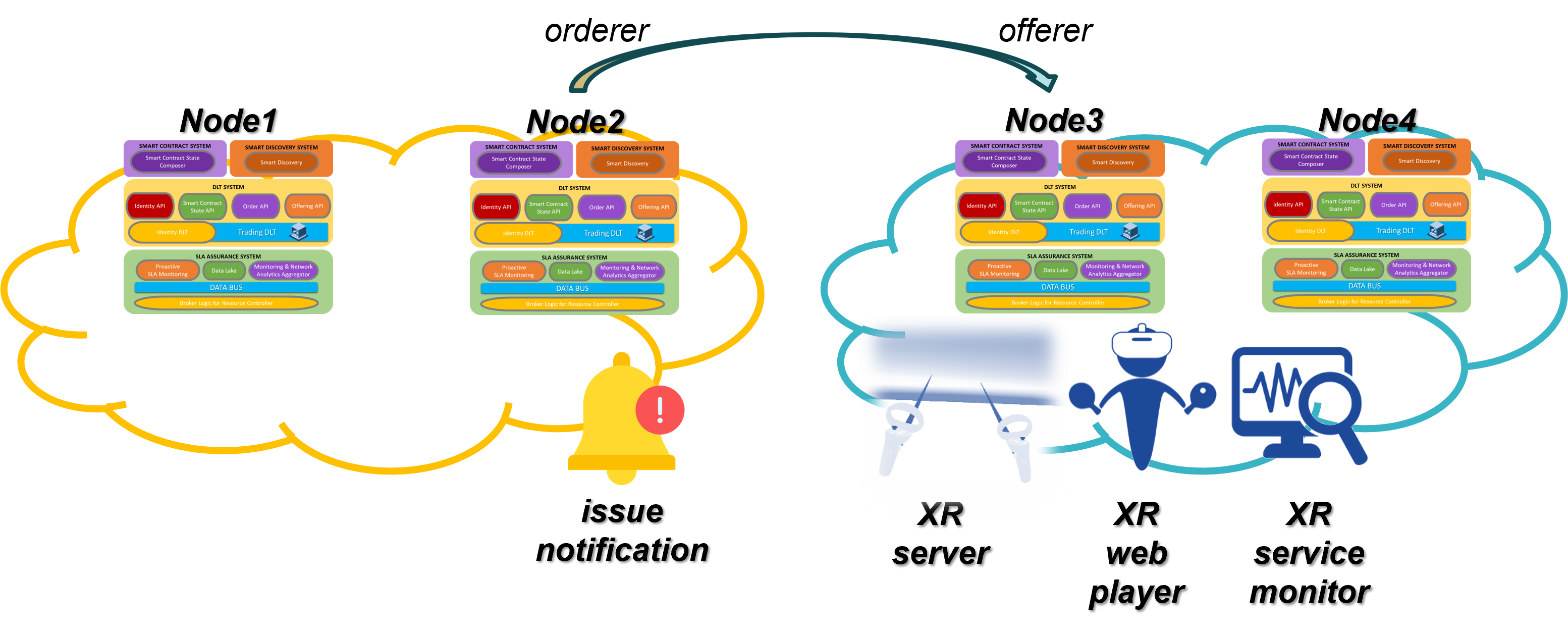}
\caption{Trading XR service through the 6GENABLERS-DLT Marketplace.}
\label{fig:testxr} % Give a unique label
\end{figure*}

The validation tests were executed across two main infrastructures, hosted by i2CAT and Vicomtech, utilizing distributed marketplace nodes. The system architecture incorporates Trading APIs, Smart Contracts, SLA Monitoring systems, and DLT frameworks, ensuring robust integration of resources and workflows across domains. The validation setup, depicted in Figure \ref{fig:testxr}, illustrates the distributed nodes’ role in enabling eXtended Reality (XR) Remote Renderer service deployment and SLA monitoring for web players consuming the service.

Each of the four use cases encapsulates specific operational workflows, validated through distinct test cases:
\begin{itemize}
    \item VNF Trading: Focused on resource acquisition for deploying virtual network functions. Validation included offer creation, discovery, order processing, and SLA adherence.
    \item Edge/Cloud Resource Trading and Provisioning: Evaluated resource trading and provisioning for deploying edge and cloud infrastructure. SLA metrics were monitored to assess the quality of service.
    \item Network Slice Trading and Provisioning: Tested trading and allocation of network slices, integrating RAN, Core, and Edge capabilities. SLA breach notifications and proactive alerts were generated during testing.
    \item Network Service Trading and Provisioning: Validated service-level transactions and multi-domain orchestration. The marketplace's ability to enforce SLAs during service delivery was critical.
\end{itemize}

The validation employed scripted automation and transaction tracking to ensure consistent and repeatable measurements.

The SLA monitoring system effectively parsed SLA terms, retrieved metrics, and issued alarms upon detecting breaches. Proactive alerts provided early warnings, enabling stakeholders to take corrective actions before formal violations occurred. SLA metrics monitored during the XR service validation included latency, throughput, and service availability.

Scalability tests demonstrated the marketplace’s ability to handle increased transaction volumes and diverse resource types across nodes. The system sustained performance under varying workloads, meeting predefined KPIs for response time and operational stability.

The DLT framework, using Hyperledger Fabric and Hyperledger Indy, provided a robust foundation for distributed transaction management. Tests confirmed the system’s resilience against node failures and its ability to maintain transaction integrity and privacy through distributed identities.

The KPI validation primarily focused on SLA adherence, transaction throughput, and latency. All KPIs were achieved within acceptable thresholds, confirming the platform’s readiness for real-world deployment. Table \ref{tab:results} shows the KPIs from the evaluation performed of platform capabilities and use cases.

\begin{table}[h]
\begin{tabular}{lcc}
\hline \verspb

\textbf{Category} & \textbf{KPI} & \textbf{Value}
\verspb \verspb \hline \verspt
\makecell{Platform \\Onboarding} & \makecell{Time required by automated \\installers to instantiate SW stack} & \makecell{\textless 7 minutes \\deployed on top of K8s cluster}  
\verspb \verspb \hline \verspt
\makecell{Platform \\Scalability} & \makecell{Linear performance degradation \\under load} & \makecell{\textless 1\% on CPU and RAM use \\when adding more nodes \\or perform/receive new transactions}
\verspb \verspb \hline \verspt
\makecell{Platform \\Interoperability} & \makecell{Administrative network domains \\and MANO frameworks} & \makecell{2 networks domains \\including i2CAT and Vicomtech \\Testbeds and different NaaS APIs}
\verspb \verspb \hline \verspt
\makecell{Platform \\Cost} & \makecell{OPEX of required infrastructure \\to host marketplace} & \makecell{20 vCPUs RAM: 128GB Disk: 120 GB \\equivalent to r6g.4xlarge Amazon EC2}
\verspb \verspb \hline \verspt
\makecell{Use Case \\Throughput} & \makecell{Transaction demands of \\simultaneous multi-party interactions} & \makecell{\textless 5 kbps \\from 3 types of TM Forum documents \\and 4 nodes in the marketplace}  
\verspb \verspb \hline \verspt
\makecell{Use Case \\Distribution} & \makecell{SLA breach notifications issued \\within target timeframes} & \makecell{\textless 500 ms \\after breach detection}
\verspb \verspb \hline
\end{tabular}
\caption{\label{tab:results}KPIs of Patform and Use Cases.}
\end{table}
% For instance:
% \begin{itemize}
%     \item SLA breach notifications were issued within target timeframes (e.g., 500 ms after breach detection).
%     \item Transaction throughput met the demands of simultaneous multi-party interactions.
%     \item Scalability tests showed linear performance degradation under load.
% \end{itemize}

The comprehensive evaluation across these dimensions validates the marketplace's capability to support dynamic 6G services, highlighting its resilience and efficiency in handling complex resource trading scenarios.

\section{Key Results and Lessons Learned}
\label{sec:lessons}
% Summary of the main project results and their implications for the telecommunications industry.
% Lessons learned from the development, testing, and validation processes.
% Discussion of remaining challenges and areas for future improvement.

The 6GENABLERS-DLT project has achieved significant milestones, delivering a distributed marketplace for resource trading tailored to the demands of 6G telecommunications infrastructure and services. The project has provided a comprehensive validation of DLT-anchored trading mechanisms, SLA monitoring systems, and a suite of APIs and smart contract tools. These contributions mark a transformative step toward decentralization, resilience, and transparency in telco resource management, with far-reaching implications for the telecommunications industry.  

The key results are:
\begin{itemize}
    \item Implementation of a Decentralized Marketplace: The project successfully deployed a distributed marketplace across multiple domains, integrating Hyperledger Fabric and Indy for trading and identity management. The marketplace facilitated resource sharing with built-in privacy, transparency, and security, enabling multi-party collaboration.  
    \item Advanced SLA Assurance Mechanisms: SLA monitoring, prediction, and alerting systems demonstrated their ability to track compliance with SLA terms and proactively mitigate risks, supporting higher reliability in 6G environments.  
    \item Smart Contract Systems for Trading: The development of a flexible smart contract framework, coupled with APIs and a developer toolkit, enabled efficient resource negotiation, order management, and lifecycle handling, supporting a variety of use cases such as VNF trading, edge/cloud resource provisioning, and network slice trading.  
    \item Operational Integration with Network Management Systems: The broker logic played a pivotal role in translating marketplace transactions into actionable configurations for underlying network management systems, ensuring seamless service delivery.  
    \item Scalability and Resilience Validated: Tests confirmed the marketplace's ability to handle increased workloads and sustain performance under stress, proving its readiness for dynamic 6G scenarios.  
\end{itemize}

The project also highlighted several challenges that informed key lessons for future development:
\begin{itemize}
    \item Cascading Changes in Data Models: Modifying data models required simultaneous updates across multiple components, including the DLT, chaincodes, APIs, and the smart contract development toolkit. This underscored the need for a more modular and adaptable design to simplify such changes.  
    \item Adaptability of Broker Logic: The lack of a standardized Network as a Service (NaaS) API necessitated extensive customization in the broker logic to ensure compatibility with various network management systems. This reinforces the importance of advancing standardization efforts in NaaS APIs.
    \item Limitations of DID Systems: Using DID systems to manage relationships between entities in the marketplace proved challenging. Metadata had to be added to smart contracts to declare relationships, highlighting the need for enhancements in DID systems to support relational data structures.
    \item Consensus Requirements in Hyperledger Indy: The platform's requirement for a minimum of four nodes for consensus limited its applicability in scenarios with fewer nodes, calling for more flexible configurations to support smaller deployments.  
    \item Lack of Standardized Metric Dictionaries: The absence of a standardized dictionary for network, infrastructure, and service metrics complicated SLA mapping, requiring custom configurations to align generic metric names with telemetry system nomenclature.  
    \item Restoration of Trading DLT Nodes: Recovering a trading DLT node after failure proved difficult due to challenges in reconnecting to preexisting data channels. This emphasized the importance of robust restoration procedures and better failover mechanisms.  
\end{itemize}

While the project achieved its objectives, several challenges remain as opportunities for future work:
\begin{itemize}
    \item Streamlining modularity across marketplace components to minimize the impact of cascading changes.  
    \item Driving standardization efforts in NaaS APIs and metric dictionaries to simplify integration and SLA mapping.  
    \item Enhancing DID systems to natively support relational data, eliminating the need for metadata workarounds.  
    \item Optimizing DLT frameworks for smaller deployments, enabling broader applicability.  
    \item Improving fault tolerance and recovery mechanisms for DLT nodes, ensuring seamless restoration in case of failures.  
\end{itemize}

These lessons and results position the 6GENABLERS Marketplace as a robust foundation for future 6G networks, providing valuable insights for evolving the platform and addressing emerging challenges in decentralized telecommunications ecosystems.

\section{Impact and Future Outlook}
\label{sec:impact}
% Contribution of 6GENABLERS-DLT to advancing 6G technologies and decentralized telco ecosystems.
% Potential for scaling and adoption by industry stakeholders.
% Alignment with broader 6G objectives, such as sustainability and innovation.

The 6GENABLERS-DLT project has made significant strides in advancing the concept of a DLT-anchored distributed telco marketplace for 6G networks, facilitating resource trading and SLA management in a decentralized environment. This section discusses the project's public dissemination efforts, contributions to 6G technologies, potential for adoption, and alignment with broader 6G objectives.  

To foster engagement with industry stakeholders, academia, and the public, the project produced several blog posts that highlight its innovations, challenges, and outcomes. These include:  
\begin{itemize}
    \item \href{https://i2cat.net/6genablers-dlt-empowering-6g-networks-with-a-dlt-anchored-smart-marketplace-for-decentralized-telco-resource-trading/}{Empowering 6G networks with DLT}: Introducing the vision and key goals of the 6GENABLERS Marketplace \cite{blogvision}.
    \item \href{https://i2cat.net/identifying-the-ideal-dlt-for-a-telco-marketplace-insights-from-the-6genablers-dlt-project/}{Evaluating DLT technologies for telco marketplaces}: Exploring the performance and suitability of different DLT technologies \cite{blogdlt}.  
    \item \href{https://i2cat.net/the-smart-contract-system-in-6genablers-dlt-the-door-to-a-telco-marketplace/}{The Smart Contract System in 6GENABLERS-DLT}: Detailing the role of smart contracts in enabling resource trading \cite{blogsc1}.  
    \item \href{https://i2cat.net/from-intent-to-action-the-role-of-smart-discovery-in-the-6genablers-marketplace/}{The role of Smart Discovery in the marketplace}: Highlighting the intelligent mechanisms for matching consumer intent with resource offers \cite{blogsd}.
    \item \href{https://i2cat.net/from-monitoring-to-prevention-how-sla-assurance-elevates-the-6genablers-marketplace/}{SLA assurance and its impact on resource trading}: Discussing proactive SLA monitoring and its transformative effects on resource trading \cite{blogsla}.  
    \item \href{https://i2cat.net/the-business-potential-of-smart-contracts/}{The business potential of Smart Contracts}: Exploring how smart contracts open new business opportunities in the telco ecosystem \cite{blogsc2}.  
\end{itemize}

Additionally, this white paper summarizing the project's key results and implications has been published in \textit{arxiv}, alongside a video demonstration showcasing the deployment of an XR service using the 6GENABLERS Marketplace, available on social media platforms.  

The project has achieved notable dissemination through high-impact scientific publications and conference presentations, including:  
\begin{itemize}
    \item Journal Publications
    \begin{itemize}
        \item "Unlocking the Path Towards Intelligent Telecom Marketplaces for Beyond 5G and 6G Networks" in IEEE Communications Magazine \cite{5g6g}.
        \item "Distributed Marketplace of 6G Networks, Computing Assets and Services under SLA Control" submitted to Journal of Network and Computer Applications.
    \end{itemize}
    \item Conference Publications
    \begin{itemize}
        \item "6GENABLERS: A Holistic Approach to Establish Pervasive Trust in 6G Networks" presented at IEEE CAMAD 2023 \cite{visionproject}.
        \item "i2Slicer: Enabling Flexible and Automated Orchestration of 5G SA End-to-End Network Slices" presented at IEEE NFV-SDN 2023 \cite{i2slicer}.
        \item "A Comprehensive Analysis of Distributed Ledger Technologies for Beyond 5G Networks" presented at MeditCom 2024 \cite{dlt}.
        \item "Distributed Identity Model for 5G Networks and Beyond Trading" presented at IEEE CAMAD 2024.
        \item "Identity Management for Frameworks Managing Northbound APIs of Future Networks" submitted to EUCNC2025.
    \end{itemize}
\end{itemize}

The project also engaged with the broader 6G research community through a joint event with the SNS project 6G-XR, fostering collaboration on 6G technologies and use cases.  

The 6GENABLERS Marketplace addresses critical challenges in 6G environments, including multi-party collaboration, resource sharing, and SLA enforcement. By leveraging DLTs, the project demonstrated the feasibility of decentralized governance, transparent resource trading, and efficient SLA management. The integration of smart contracts and intelligent discovery systems enhances operational efficiency and creates new opportunities for innovation in the telecommunications industry.  

The modular design and scalable architecture of the 6GENABLERS Marketplace make it highly adaptable to diverse telco ecosystems. By addressing key operational challenges, such as SLA monitoring, smart discovery, and decentralized identity management, the project offers practical solutions that can be adopted by network operators, service providers, and infrastructure vendors. Its compatibility with TM Forum specifications and NaaS principles further supports interoperability, paving the way for industry-wide adoption.  

The project aligns with the core goals of 6G, including sustainability, resilience, and innovation. By enabling efficient resource sharing and reducing the need for centralized infrastructure, the 6GENABLERS Marketplace contributes to energy efficiency and environmental sustainability. Its decentralized architecture enhances network resilience, ensuring robust performance even in dynamic scenarios. Finally, by fostering collaboration and facilitating new business models, the project drives innovation across the telecommunications value chain.  

Through its comprehensive approach, 6GENABLERS-DLT lays a strong foundation for the evolution of decentralized telco ecosystems, empowering stakeholders to harness the full potential of 6G networks.

\section{Conclusions}
\label{sec:conclusions}
%\hl{Be positive}
% Recap of the project's significance and the benefits of the developed solutions.
% Call to action for industry stakeholders to leverage the marketplace for future collaboration and innovation.

The 6GENABLERS-DLT project has demonstrated the transformative potential of a DLT-anchored distributed telco marketplace for 6G networks, addressing critical challenges in resource trading, SLA enforcement, and multi-party collaboration. By leveraging cutting-edge technologies such as distributed ledger systems, smart contracts, proactive SLA monitoring, and intelligent discovery mechanisms, the project has established a robust framework for decentralized telco ecosystems.  

The developed marketplace introduces unprecedented transparency, flexibility, and efficiency, allowing stakeholders to seamlessly discover, trade, and manage 6G resources across a wide range of use cases. From VNF trading to network slice provisioning, the 6GENABLERS Marketplace showcases how decentralized architectures can streamline operations, enhance trust, and foster innovation in a rapidly evolving telecommunications landscape. The modular design, adherence to TM Forum standards, and scalable architecture ensure that the solutions developed are not only technically advanced but also practically viable for real-world adoption.  

As the telecommunications industry embraces 6G technologies, the outcomes of the 6GENABLERS-DLT project serve as a call to action for industry stakeholders. Network operators, service providers, infrastructure vendors, and other ecosystem participants are encouraged to adopt and build upon the solutions developed in this project. By integrating these tools into their operations, stakeholders can unlock new opportunities for collaboration, develop innovative business models, and drive the next generation of connectivity services.  

The vision of a decentralized, resilient, and efficient telco marketplace is now within reach. By leveraging the foundational work laid by the 6GENABLERS-DLT project, industry players can transform the way resources and services are exchanged, ensuring the telecommunications sector remains at the forefront of technological advancement and societal progress.

\section{Acknowledgement}
This work was supported by the Spanish Ministry of Economic Affairs and Digital Transformation and the European Union – NextGenerationEU, in the framework of the Recovery Plan, Transformation and Resilience (PRTR) (Call UNICO I+D 5G 2021, ref. number TSI-063000-2021-12 - 6GENABLERS-DLT).

\pagebreak

\section{List of Contributors}

Table \ref{tab:contribs} shows the list of contributors to this document.

\begin{table}[h]
\begin{tabular}{cl}
\hline \verspb

\textbf{Affiliation} & \multicolumn{1}{c}{\textbf{Name}}

\verspb \verspb \hline \verspt

i2CAT       & \begin{tabular}[c]{@{}l@{}}Shuaib Siddiqui, Alfonso Egio, Mario Montagud, Sergio Giménez, \\Adriana Fernández Fernández\end{tabular}

\verspb \verspb \hline \verspt

VICOMTECH   & \begin{tabular}[c]{@{}l@{}}Juncal Uriol, Felipe Mogollon, Breno da Costa, Emma O'Brien, Roberto Viola, \\Angel Martin\end{tabular}

\verspb \verspb \hline \verspt

ATOS       & \begin{tabular}[c]{@{}l@{}}Daniel Ruiz, Sonia Castro, Guillermo Gomez\end{tabular}

\verspb \verspb \hline

\end{tabular}
\caption{\label{tab:contribs}List of Contributors.}
\end{table}

% \section{Backup}

% To view tutorials, user guides, and further documentation, please visit our \href{https://www.overleaf.com/learn}{help library}, or head to our plans page to \href{https://www.overleaf.com/user/subscription/plans}{choose your plan}.

% See the code for Figure \ref{fig:frog} in this section for an example.

% \begin{figure}
% \centering
% \includegraphics[width=0.3\textwidth]{frog.jpg}
% \caption{\label{fig:frog}This frog was uploaded via the file-tree menu.}
% \end{figure}

% You can make lists with automatic numbering \dots

% \begin{enumerate}
% \item Like this,
% \item and like this.
% \end{enumerate}
% \dots or bullet points \dots
% \begin{itemize}
% \item Like this,
% \item and like this.
% \end{itemize}

% Let $X_1, X_2, \ldots, X_n$ be a sequence of independent and identically distributed random variables with $\text{E}[X_i] = \mu$ and $\text{Var}[X_i] = \sigma^2 < \infty$, and let
% \[S_n = \frac{X_1 + X_2 + \cdots + X_n}{n}
%       = \frac{1}{n}\sum_{i}^{n} X_i\]
% denote their mean. Then as $n$ approaches infinity, the random variables $\sqrt{n}(S_n - \mu)$ converge in distribution to a normal $\mathcal{N}(0, \sigma^2)$.

% You can then cite entries from it, like this: \cite{greenwade93}. Just remember to specify a bibliography style, as well as the filename of the \verb|.bib|. Contact form at \url{https://www.overleaf.com/contact}.

\pagebreak

\bibliographystyle{alpha}
\bibliography{main}

\end{document}